\documentclass[twocolumn]{aastex631}
\usepackage{stfloats}
\usepackage{float}
\usepackage{graphicx}
\usepackage{natbib}
\usepackage{hyperref}
\usepackage{amsmath}
\usepackage{autobreak}
\usepackage{multirow}
\usepackage{makecell}
\usepackage{color}
\usepackage{bm}
\usepackage{threeparttable}
\usepackage[figuresright]{rotating}
\usepackage[mathscr]{euscript}

\usepackage[utf8]{inputenc}
\usepackage[T1]{fontenc}
\usepackage{amsmath}
\usepackage{makecell}
\usepackage{subfigure}

\begin{document}

\title{Constraining the progenitor of the nearby Type~II-P SN~2024ggi with environmental analysis}

\correspondingauthor{Ning-Chen Sun}
\email{sunnc@ucas.ac.cn}

\author{Xinyi Hong}
\affiliation{School of Astronomy and Space Science, University of Chinese Academy of Sciences, Beijing 100049, People's Republic of China}
\affiliation{National Astronomical Observatories, Chinese Academy of Sciences, Beijing 100101, China}

\author{Ning-Chen Sun}
\affiliation{School of Astronomy and Space Science, University of Chinese Academy of Sciences, Beijing 100049, People's Republic of China}
\affiliation{National Astronomical Observatories, Chinese Academy of Sciences, Beijing 100101, China}
\affiliation{Institute for Frontiers in Astronomy and Astrophysics, Beijing Normal University, Beijing, 102206, People's Republic of China}

\author{Zexi Niu}
\affiliation{School of Astronomy and Space Science, University of Chinese Academy of Sciences, Beijing 100049, People's Republic of China}
\affiliation{National Astronomical Observatories, Chinese Academy of Sciences, Beijing 100101, China}

\author{Junjie Wu}
\affiliation{National Astronomical Observatories, Chinese Academy of Sciences, Beijing 100101, China}
\affiliation{School of Astronomy and Space Science, University of Chinese Academy of Sciences, Beijing 100049, People's Republic of China}

\author{Qiang Xi}
\affiliation{School of Astronomy and Space Science, University of Chinese Academy of Sciences, Beijing 100049, People's Republic of China}
\affiliation{National Astronomical Observatories, Chinese Academy of Sciences, Beijing 100101, China}

\author{Jifeng Liu}
\affiliation{National Astronomical Observatories, Chinese Academy of Sciences, Beijing 100101, China}
\affiliation{School of Astronomy and Space Science, University of Chinese Academy of Sciences, Beijing 100049, People's Republic of China}
\affiliation{Institute for Frontiers in Astronomy and Astrophysics, Beijing Normal University, Beijing, 102206, People's Republic of China}
\affiliation{New Cornerstone Science Laboratory, National Astronomical Observatories, Chinese Academy of Sciences, Beijing 100012, People's Republic of China}

\begin{abstract}

The progenitors of Type~II-P supernovae (SN) have been confirmed to be red supergiants. However, the upper mass limit of the directly probed progenitors is much lower than that predicted by current theories, and the accurate determination of the progenitor masses is key to understand the final fate of massive stars. Located at a distance of only 6.72~Mpc, the Type II-P SN~2024ggi is one of the closest SN in the last decade. Previous studies have analyzed its progenitor by direct detection, but the derived progenitor mass may be influenced by the very uncertain circumstellar extinction and pulsational brightness variability. In this work, we try to constrain the progenitor mass with an environmental analysis based on images from the Hubble Space Telescope. We found that stars in the progenitor environment have a uniform spatial distribution without significant clumpiness, and we derived the star formation history of the environment with a hierarchical Bayesian method. The progenitor is associated with the youngest population in the SN environment with an age of log($t$/yr) = 7.41 (i.e. 25.7~Myr), which corresponds to an initial mass of $10.2^{+0.06}_{-0.09}$~$M_\odot$. Our work provides an independent measurement of the progenitor mass, which is not affected by circumstellar extinction and pulsational brightness variability. 

\end{abstract}

\section{Introduction} \label{sec:intro}

Core-collapse supernovae (CCSNe) are the explosion of dying massive stars of  $>$8~$M_\odot$. As the most common type of CCSNe, Type~II-P is defined by the presence of strong H lines in the spectra and a plateau phase in the light curve. With $\sim$20 progenitor detections on pre-explosion images, it has been confirmed that red supergiants (RSGs) are the direct progenitors of Type~II-P supernovae (SNe) \citep{2009Smartt, 2017VanDyk, 2023Niu} .

However, there are still some major problems in the study of Type~II-P SNe progenitors. For example, the upper mass limit of directly probed progenitors (16~$\pm$~1.5~$M_\odot$) is significantly lower than that predicted by current stellar evolutionary theories (25-30~$M_\odot$; \citealp{2009Smartt}; i.e. the "RSG problem"). It has been proposed that RSGs of $>$17~$M_\odot$ will collapse directly into a black hole \citep{2017O'Connor} or explode as other types of SNe (e.g. II-L, IIn, IIb; \citealp{2011Leonard, 2013Groh}). It may also be due to circumstellar extinction, which could lead to significant underestimation of the progenitor masses (\citealp{2017Davies}). It is also worth noting that some RSG progenitors have pulsational brightness variability, which may also affect the mass estimate significantly (\citealp{2023Niu, 2024Beasor}). 
In addition, \citet{2024Beasor} finds that uncertain bolometric correction can cause error in the mass estimation as well.

Discovered on April 11, 2024 (JD=2460411.64069) \citep{2024Tonry}, SN~2024ggi is a Type II-P SN \citep{2024ChenT} exploded in the nearby galaxy NGC~3621 at a distance of only 6.72 Mpc \citep{2013Tully}. Several research groups have made extensive multi-band photometric and spectroscopic follow-up observations. In particular, \citet{2024Zhang} detected the shock breakout signal from the early-time, hour-to-day cadence spectroscopy. By probing the flash-ionzied emission lines, researchers had found that SN 2024ggi is surrounded by a dense, optically thick  circumstellar material (CSM) \citep{2024Jacobson-Gal, 2024Pessi, 2024Shrestha, 2024ChenT}. \citet{2024Soker} explained its CSM with an effervescent model and discovered that the explosion is caused by the jittering jets explosion mechanism.

\citet{2024Xiang} analyzed the pre-explosion images of SN 2024ggi from the Hubble Space Telescope (HST), Spitzer Space Telescope and VISTA Hemisphere Survey (VHS). They found a progenitor with a very red color, indicating very high circumstellar extinction, and with pulsational brightness variability that can reach 0.99~$\pm$~0.03 mag in the F814W band between two observational epochs. By carefully analyzing its spectral energy distribution (SED), they derived an initial mass of 13~$M_\odot$. \citet{2024Chen} also studied the progenitor based on the images of the Dark Energy Camera Legacy Survey (DECaLS), with which they derived an initial mass of 14--17~$M_\odot$.

Analysis of SN environment also serves as a powerful way to constrain the progenitor properties (\citealp{2009Gogarten, 2011Murphy, 2018Díaz-Rodríguez, 2019Auchettl}; \citealp{2020Sun}, \citeyear{2021Sun}, \citeyear{2022Sun}, \citeyear{Sun2023a}, \citeyear{Sun2023b}) since massive stars often form in groups and stars within each group have similar ages and metallicities. This method is not affected by the possible influence of circumstellar extinction and pulsational brightness variability of the progenitor itself. Therefore, it can provide an independent measurement of progenitor mass.

This paper reports a detailed analysis of the environment of SN~2024ggi to obtain the accurate initial mass of its progenitor. Throughout this paper, we adopt a distance module of $\mu$ = 29.14 mag (\citealp{2013Tully}), a Milky Way extinction of $E(B-V)_{\rm{MW}}$ = 0.072 mag (\citealp{2011Schlafly}), and a standard extinction law with $R_V$ = 3.1 (\citealp{1989Cardelli}).

\section{Data}

Data used in this work are obtained by HST. We used the pre-explosion images conducted by Wide Field Camera 3 (WFC3) Ultraviolet Visible (UVIS) channel, and the Advanced Camera for Surveys (ACS) Wide Field Channel (WFC). The observations by ACS/WFC were in February 2003 and WFC3/UVIS in December 2019 and May 2023, covering a wide wavelength range from the F275W band to the F814W band. The observations in the F814W filter were composed of four separate dithered exposures, and those in the other filters were composed of three exposures. 
More details such as proposal ID and total exposure time for each filter are presented in Table~\ref{tab:data}. We retrieved the data from Mikulski Archive for Space Telescopes (MAST)\footnote{\url{https://mast.stsci.edu/search/ui/\#/hst}}.

For all the observations, we used \textsc{tweakreg} to align images of all filters. Then, we combined the exposures by filter with the \textsc{astrodrizzle} to remove cosmic-ray contamination. In this process, we set \texttt{driz\_cr\_grow = 3} for more efficient removal of cosmic rays, while other drizzle parameters were kept unchanged as in the standard HST calibration pipeline. Photometry was performed with \textsc{dolphot} \citep{2000Dolphin, 2016Dolphin}. We use \texttt{FitSky = 2} and \texttt{RAper = 3}, which are recommended by the user manual for crowded regions. As all the images have already been corrected for charge transfer inefficiency, there is no need to do additional correction for both ACS and WFC3 observations (i.e. we used \texttt{ACSuseCTE = 0, WFC3useCTE = 0}). In addition, we set \texttt{Force1 = 1} (force all objects to be fitted as stars), \texttt{ApCor = 1} (turn on aperture correction), and \texttt{UseWCS = 1} (use WCS header information for alignment). All other parameters are the same as recommended by the \textsc{dolphot} user manual.

\begin{table}[H]
\centering

\caption{HST observation of SN2024ggi. \label{tab:data}}
    \begin{tabular}{ccccccc}
    \hline
    \hline
    
     Proposal & Date &  Instrument & Filter & Exposure  \\
     ID & & & & Time (s) \\
    
    \hline
    9492 & 2003-02-03 & ACS/WFC   & F435W & 1080 \\
           & 2003-02-03 & ACS/WFC   & F555W & 1080 \\
           & 2003-02-03 & ACS/WFC   & F814W & 1440 \\
           
    \hline
    15654 & 2019-12-13 & WFC3/UVIS & F275W & 2322 \\
           & 2019-12-13 & WFC3/UVIS & F336W & 2282 \\
          
    \hline
    17189 & 2023-05-19 & WFC3/UVIS & F300X & 2442 \\

    \hline
 \end{tabular}

     \begin{tablenotes}
\item[1] PIs: 9492: Bresolin; 15654: Lee; 17189: Thilker.
      \end{tablenotes}

\end{table}

\section{Analysis of the SN environment} \label{sec:env}

\begin{figure*}[htbp]
    \centering
    \includegraphics[width=0.9\linewidth]{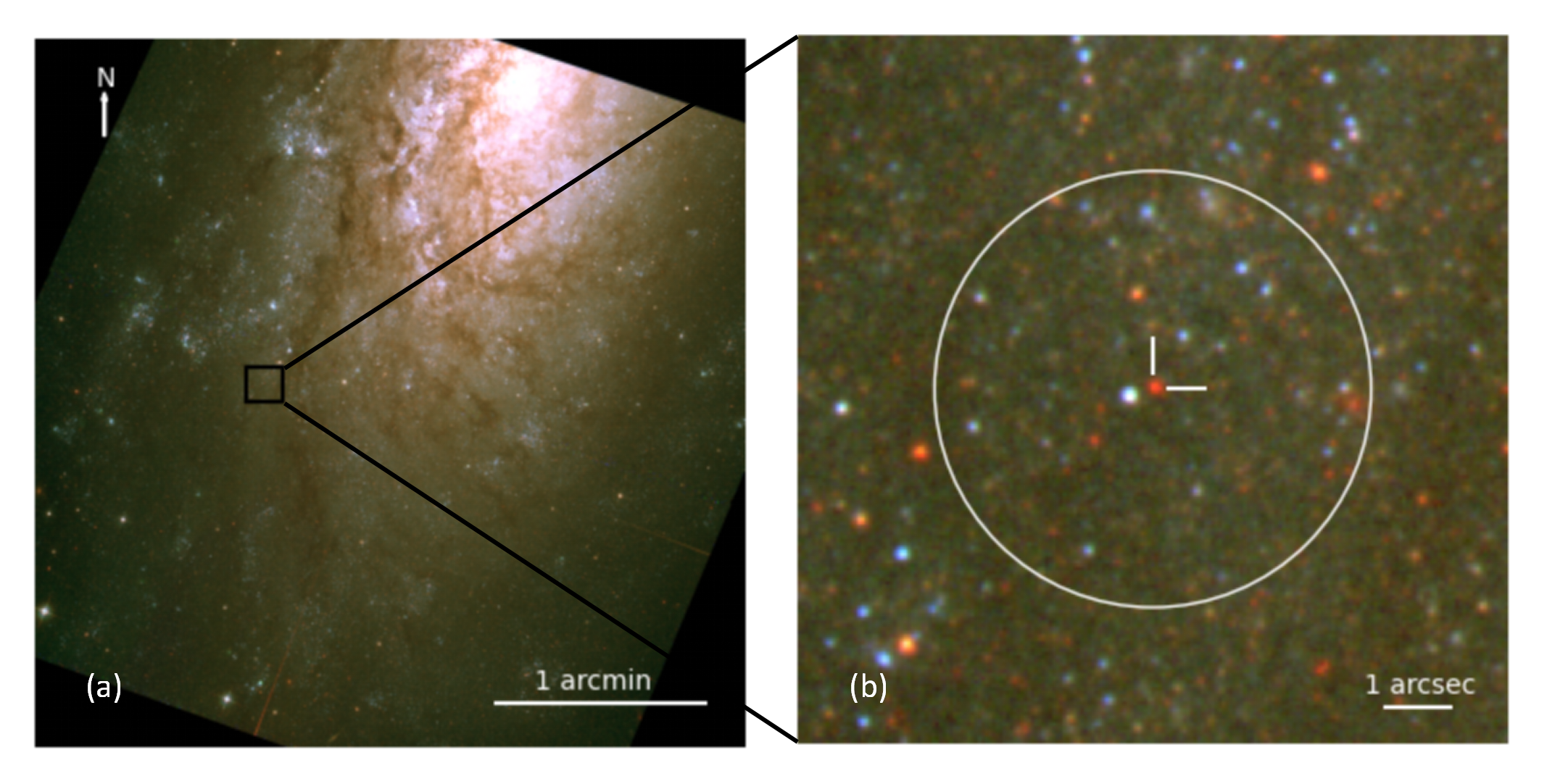}
    \caption{F435W/F555W/F814W three-colour composite image of the environment of SN~2024ggi by HST observations. The white circle in panel (b) has a radius of 100~pc, within which we studied the star formation history in the local environment.}
    \label{fig:observation}
\end{figure*}

The SN\,2024ggi is located in the nearby galaxy NGC 3621 (Figure~\ref{fig:observation}a). 
It is a spiral galaxy with a few dark lanes caused by dust in the disk. SN\,2024ggi is located in the outskirt of the host galaxy and to the east of a dark lane. The local environment within 100~pc (Figure~\ref{fig:observation}b) is characterized by a relatively uniform stellar distribution without significant clumpiness.

In order to perform a quantitative analysis of the star-forming history in the SN environment, we defined a circular region with a radius of 100 pc around the SN (see the white circle in Figure~\ref{fig:observation}b). 
We selected good stars based on the following criteria:
\begin{enumerate}
    \item Type of source, \texttt{obj = 1}
    \item Signal-to-noise ratio, \texttt{snr > 5}
    \item Source sharpness, \texttt{|shp| < 0.5}
    \item Source crowding, \texttt{crd < 2}
    \item Photometry quality flag, \texttt{qfg < 3}
\end{enumerate}
We finally selected a catalog of 238 sources (excluding the progenitor candidate) for further analysis. 

\begin{figure*}[htbp]
    \centering
    \includegraphics[width=0.9\linewidth]{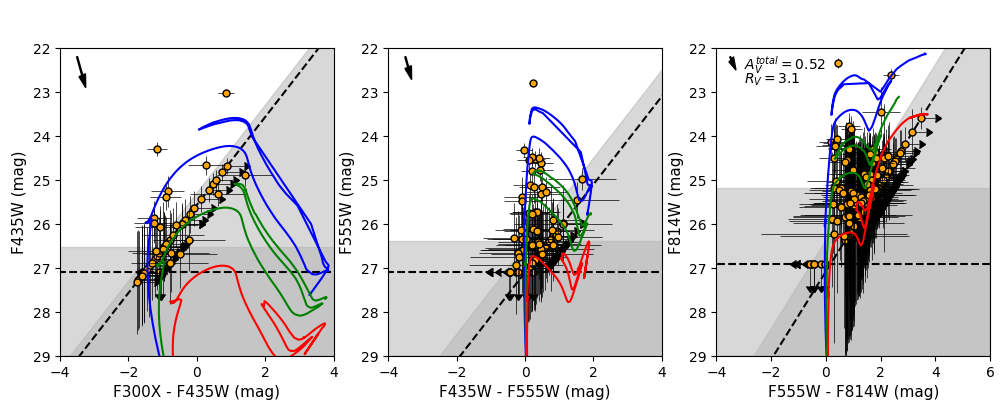}
    \caption{CMDs of all stellar sources in the environment of SN 2024ggi. In all panels, the blue, green and red isochrones correspond to stellar populations with ages of log($t$/yr) = $7.41$, $7.75$, $8.09$, respectively. The arrows in the upper left corner are reddening vectors for $R_V$ = 3.1 and total extinctions $A_V^{\rm{total}}$ = 0.52, which corresponds to a best-fitting host extinction $A_V^{\rm{host}}$ = $0.30$ and a Milky Way reddening of $E(B-V)_{\rm{MW}}$ = 0.072 mag.}
    \label{fig:CMD}
\end{figure*}

We conducted artificial star tests to estimate the detection limits for each filter. It is reasonable to assume that the detection limit does not have spatial variations in the SN environment since stars have a very uniform spatial distribution. So in each filter, the artificial stars with different magnitudes are randomly placed in the region. An artificial star is considered to be successfully recovered when its \textsc{dolphot} results meet all the above criteria. Thus, for each magnitude, we obtain a detection probability in the region. We consider the magnitude at which the detection probability falls to 50\% as the detection limit. We also use artificial star tests to calculate a scaling factor of the reported photometric uncertainty to account for the additional error introduced by source crowding. 
Figure~\ref{fig:CMD} shows the colour-magnitude diagrams (CMDs) drawn with the catalog.

Based on this catalog, we use a hierarchical Bayesian method (see \citealp{2016Maund} and \citealp{2021Sun} for detailed description) to fit the stars with model stellar populations (based on \textsc{parsec}\footnote{\url{http://stev.oapd.inaf.it/cgi-bin/cmd}} stellar isochrones, \citealp{2012Bressan}). In this process, we assume a \citet{1955Salpeter} initial mass function, a 50\% binary fraction, and a flat distribution of primary-to-secondary mass ratio; all the binaries are considered as non-interacting systems; the metallicity is assumed to be solar (consistent with \citealt{2024Xiang}).

It has been known that a prolonged star formation can be considered as the mixture of multiple short-term star formation \citep{2013Walmswell, 2016Maund}. Therefore, we assume that each model population have a Gaussian distribution with a small standard deviation of 0.05 dex in stellar log-ages, and the mean log-ages all have flat priors between 6.60 to 8.60. Meanwhile, as the stars in the environment have a very uniform spatial distribution, the mean host stellar extinctions of all model populations can be considered as the same, for which we adopted a Gaussian prior based on the measurement of \citet{2024Zhang} ($E(B-V)_{\rm{host}}$ = 0.113~$\pm$~0.014 mag). In addition, we assume that stars within each model population may have a small extinction dispersion of $\sigma_{A_V}$ = 0.05 to account for the possible star-to-star extinction differences.
It has been found that altering the value of $R_V$ has only a very small effect on the derived ages \citep{2022Sun}, so we used the standard extinction law with $R_V$ = 3.1 \citep{1989Cardelli}.

\begin{figure*}[htbp]
    \centering
    \includegraphics[width=0.8\linewidth]{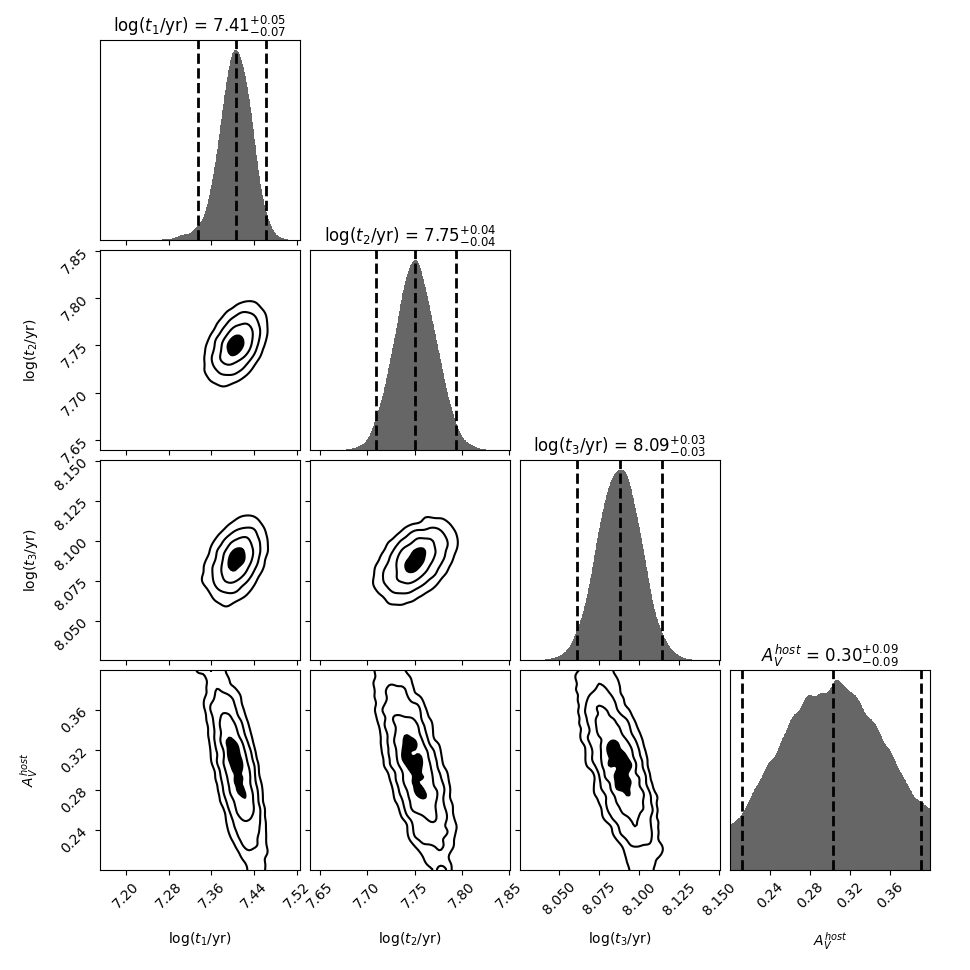}
    \caption{Posterior probability distributions of mean stellar log-ages of model stellar populations and the host stellar extinction. The numbers on top of and the dashed lines in each histogram show the median value with the 95\% (2-sigma) credible interval of the marginalised posterior probability. The contours in the other panels contain 0.5, 1, 1.5, 2-sigma marginalised posterior probability from inside to outside.}
    \label{fig:bayesfit}
\end{figure*}

We find that the stars in the region can be fitted with three model stellar populations. Figure~\ref{fig:bayesfit} shows the corner plot of the posterior probability distributions. The best fitting log-ages are log($t$/yr) = $7.41^{+0.05}_{-0.07}$, $7.75^{+0.04}_{-0.04}$, and $8.09^{+0.03}_{-0.03}$ (2-sigma uncertainty), and the corresponding stellar isochrones are presented in the CMDs (see Figure~\ref{fig:CMD}).
 
On the CMDs there are several sources that are much brighter than the others in the SN environment. While they may be single stars or non-interacting binaries as considered in our model, they may also be multiple stars, star clusters, or rejuvenated stars and merger products from interacting binaries. These objects were not considered in our model stellar populations. To assess whether these objects influence our result, we repeated the above analysis by excluding those sources. The results are almost unchanged. Therefore, our result is not affected by these very bright sources.

According to the \textsc{parsec} stellar evolutionary tracks and isochrones, the mean log-ages of the three model stellar populations correspond to the lifetimes of (single) stars of initial masses of 10.2$M_\odot$, 6.8$M_\odot$, 4.9$M_\odot$. The latter two are not massive enough to undergo CCSNe. Thus, we conclude that SN 2024ggi should arise from the youngest population and its progenitor has an initial mass of $10.2^{+0.06}_{-0.09}$~$M_\odot$ (2-sigma uncertainty). 
\citet{2024ChenT} has found that SN~2024ggi has a regular explosion energy (2~$\times$~$10^{51}$~erg). As our estimated mass is a relatively low mass compared to the theoretical progenitor mass range, it indicates that even a 10$M_\odot$ progenitor can explode with a regular explosion energy (rather than a low explosion energy). 

Based on pre-explosion images, \citet{2024Xiang} and \citet{2024Chen} have tried to constrain the initial mass of the progenitor of SN\,2024ggi. \citet{2024Xiang} obtained an initial mass of 13~$M_\odot$ with HST, Spitzer, and VHS images, and \citet{2024Chen} derived a value of 14--17~$M_\odot$ with images from the DECaLS. The red color and the infrared excess indicate a large amount of circumstellar dust for the progenitor. The multi-epoch Spitzer observations reveal a pulsation-like brightness variability in the [3.6$\mu$m] and [4.5$\mu$m] filters. In the F814W band, the progenitor brightness varies by almost 1~mag between two different epochs of HST observations, which is relatively high compared with the RSGs in M31 \citep{2024Beasor}.

Our estimate of the progenitor mass is much lower than the results of \citet{2024Xiang} and \citet{2024Chen}. It may be due to the circumstellar dust, which can influence the apparent optical brightness and infrared excess, leading to uncertainties in the derived initial mass. It may also be due to the pulsational brightness variability. For example, \citet{2024Xiang} found an almost 1 mag difference between two epochs in the F814W band; such a difference in the bolometric luminosity corresponds to a large initial mass difference of at least 4~$M_\odot$. Although the Spitzer observations have enough epochs to sample the light curves, observations in the other filters (J, Ks, F814W, F555W, F435W) are too sparsely sampled to constrain the pulsational amplitude. On the other hand, our mass estimate is based on the assumption that the SN progenitor is associated with the youngest population in the SN environment. We cannot exclude the possibility that the progenitor could be associated with an even younger star-forming event, which occurred at a very low level and eludes our detection.

\section{Summary}\label{sec:dis}

SN~2024ggi is a nearby Type~II-P SN located in NGC 3621. 
In this paper, we give a detailed analysis of its environment based on HST observations to study its progenitor properties. The environment of SN~2024ggi has a relatively uniform stellar distribution without significant clumpiness. Then, with a hierarchical Bayesian method, we fit the star formation history of the SN environment with model stellar populations. We found three stellar populations with the stellar ages derived. We concluded that SN~2024ggi is from the youngest population with an age of log($t$/yr) = 7.41 (i.e. 25.7~Myr) as the other two populations are too old to undergo CCSNe. Therefore, the progenitor would have an initial mass of $10.2^{+0.06}_{-0.09}$~$M_\odot$ (2-sigma uncertainty). This mass is significantly lower than previous values based on direct progenitor detection. With the analysis of the SN environment, our work serves an independent measurement of the progenitor properties, which is not affected by circumstellar extinction and pulsational brightness variability.

\section*{acknowledgments}
This work is supported by the Strategic Priority Research Program of the Chinese Academy of Sciences, Grant No. XDB0550300.
NCS’s research is funded by the NSFC grants No. 12303051 and No. 12261141690. ZXN acknowledges support from the NSFC through grant No. 12303039. JFL acknowledges support from the NSFC through grants No. 11988101 and No. 11933004 and from the New Cornerstone Science Foundation through the New Cornerstone Investigator Program and the XPLORER PRIZE.

\vspace{5mm}

\bibliography{sample631}{}
\bibliographystyle{aasjournal}

\end{document}